\documentstyle[12pt]{article}

\def\beq{\begin{equation}}
\def\eeq{\end{equation}}
\def\lesim{{\buildrel <\over\sim}}
\def\gesim{{\buildrel >\over\sim}}

\begin{document}
\pagestyle{empty}
$\ $
\vskip 2 truecm

\centerline{\bf EFFECTS OF FRICTION ON COSMIC STRINGS}
\vskip 1 truecm
\centerline{Jaume Garriga$^{1}$ and Maria Sakellariadou$^{2}$}
\vskip .5 truecm
\centerline{\em $^1$Tufts Institute of Cosmology,}
\centerline{\em Department of Physics and Astronomy,
              Tufts University, Medford, MA 02155}
\vskip .3 truecm
\centerline{\em $^2$ Laboratoire de Mod\`eles de Physique Math\'ematique,}
\centerline{\em Facult\'e des Sciences, Parc de Grandmont, 37200 Tours, France}
\vskip 2.5 truecm

\begin{abstract}

We study the evolution of cosmic strings taking into account the frictional
force due to the surrounding radiation. We consider small perturbations
on straight strings, oscillation of circular loops and
small perturbations on circular loops. For straight strings,
friction exponentially suppresses
perturbations whose co-moving scale crosses the horizon before cosmological
time $t_*\sim \mu^{-2}$ (in Planck units), where $\mu$ is the string tension.
Loops with size much smaller than $t_*$ will be
approximately circular at the time when they start the relativistic
collapse.
We investigate the possibility that such loops will form black
holes. We find that the number of black holes which are formed through this
process is well bellow present observational limits, so this does not give
any lower or upper bounds on $\mu$. We also consider the case of
straight strings attached to walls and circular holes that can
spontaneously nucleate on metastable domain walls.

\end{abstract}

\clearpage
\pagestyle{plain}
\section{Introduction}
\label{introduction}

Cosmic strings are topological defects that may have formed during
phase transitions in the early universe
(see e.g. \cite{vi85}). Their
properties and observational consequences, especially in connection with
their possible role in the formation of large scale structure in
the universe, have been extensively studied during the past
decade (see e.g. \cite{GHV90}).

Cosmic strings of mass per unit length $\mu$ would have formed
at cosmological time of order $t_0\sim (G\mu)^{-1} t_{Pl}$
($G$ is Newton's constant and $t_{pl}$ is the Planck time). It is
well known that immediately after the phase
transition the dynamics of strings would be dominated by the force of
friction \cite{ev81,vi91}.
This force is due to the scattering of thermal particles off
the string.
Friction would dominate the dynamics until a time of
order $t_*\sim (G\mu)^{-2}t_{Pl}$. In most of the investigations about
cosmic strings, the effects of friction have been neglected. The
reason is that if cosmic strings have to play a role in galaxy
formation, then they have to form near the grand unification scale. In that
case their mass per unit length is of order $G\mu\sim 10^{-6}$ and
friction is important only for a very short period of time.

However, if strings have formed at later phase transitions, say, closer to
the electroweak scale, their dynamics would be dominated by friction
through most of the thermal history of the universe.
It is therefore of some interest to study the evolution of cosmic strings with
friction in a quantitative way.
The relativistic equation of motion for
strings with friction was given by Vilenkin \cite{vi91}. The main purpose
of this paper is to solve this equation in a few simple cases, which
should be representative of more complicated situations.

The plan of the paper is the following. In Section 2 we review the results
of Ref.\cite{vi91}, in order to fix the notation. In Section 3 we study
linearized perturbations on an infinite straight string. In Section 4 we
consider the dynamics of oscillating circular loops.
The evolution of linearized perturbations on circular loops
is discussed in Section 5.

It is known that an exactly circular loop would form a black
hole when it collapses under its tension \cite{vi81,ha90}. Since friction
tends to erase perturbations whose co-moving scale crosses the horizon
before $t_*$, loops smaller than $t_*$ will tend to be approximately
circular. In Section 6 we study the possibility that such loops form
black holes.
Note that these black holes can have masses of order $M\sim \mu
t_*$, which can be considerably large if $G\mu$ is sufficiently small.
We discuss possible observational consequences of these black holes.

Finally, in Section 7, we consider perturbations on strings which are attached
to domain walls, and in particular, to circular holes which can
spontaneously nucleate in a metastable wall. Our conclusions are
summarized in Section 8.

\section{Strings with friction}

In this section we summarize the results of Ref. \cite{vi91}, where the
equation
of motion for a string with friction was found.
A cosmic string can be represented as a two dimensional worldsheet in
spacetime $x^{\mu}=x^{\mu}(\xi^a)$, where $\xi^a$, ($a=0,1$) are two arbitrary
parameters on the worldsheet
and $x^{\mu}$ are spacetime coordinates ($\mu=0,...,4$).
In flat space, the equation of motion for the
strings is $\mu\Box x^{\nu}=0$, where $\Box$ denotes the d'Alembertian on
the worldsheet.
When one includes the effects of curvature of the spacetime and a
force of friction $F^{\nu}$, the equation of motion reads \cite{vi91}
\beq
\mu\left[-\Box
x^{\nu}+\Gamma^{\nu}_{\sigma\tau}x^{\sigma}_{,a}x^{\tau,a}\right]=
F^{\nu}(u^{\lambda}_{\bot},T,\sigma)\label{1}
\eeq
Here $\Gamma^{\nu}_{\sigma\tau}$ are the four dimensional Christoffel
symbols (greek indices run over spacetime coordinates). The force of
friction $F^{\nu}$ will depend on the temperature of the surrounding
matter, $T$, the velocity of the fluid transverse to the worldsheet
$u^{\nu}_{\bot}\equiv u^{\nu}-x^{\nu}_{,a}x^{\sigma,a}u_{\sigma}$, and the
type of interaction between the particles and the string, which we
symbolically represent by $\sigma$. Vilenkin \cite{vi91}
found the form of $F^{\nu}$
for the case when friction is dominated by Aharonov-Bohm scattering of
charged particles with the pure gauge field outside the string \cite{alwi89}
\beq
F^{\nu}=\beta T^3u^{\nu}_{\bot}.\label{2}
\eeq
The numerical coefficient $\beta$ is given by
$$\beta=2\pi^{-2}\xi(3)\sum_ab_a\sin^2(\pi\nu_a)$$
where the summation is over all effectively massles degrees of freedom
(i.e. mass $m<<T$) $b_a=1$ for bosons, $b_a=3/4$ for fermions and $\nu_a$
is the phase change experienced by a particle as it is transported around
the string.

For the case of a
Friedman-Robertson-Walker (FRW) universe,
\beq
ds^2=a^2(\tau)[-d\tau^2+d\vec x^2],\label{frw}
\eeq
with the four-velocity of the fluid given by $u^{\nu}=(a^{-1},0,0,0)$, and
choosing the gauge $\xi^0=\tau,\dot{\vec x}\cdot\vec x'=0$, Eq.(\ref{1})
reduces to
\beq
\ddot{\vec x}-\epsilon^{-1}\left({\vec x'\over\epsilon}\right)'+
[2a(H+h)]
(1-\dot{\vec x}^2)\dot{\vec x}=0,\label{eom1}
\eeq
\beq
\dot\epsilon+[2a(H+h)]
\dot{\vec x}^2\epsilon=0. \label{eom2}
\eeq
Here $\dot{\ }\equiv d/d\tau$, $'\equiv d/d\xi^1$, $H\equiv\dot a/a^2$ is the
Hubble parameter and
$$
h\equiv{\beta T^3\over2\mu}.
$$
The quantity $\epsilon\equiv[\vec x'^2/(1-\dot{\vec x}^2)]^{1/2}$ is related
to the energy of the string \cite{vi85}
\beq
E=\mu a(\tau)\int d\xi^1\epsilon.\label{energy}
\eeq
[Using the definition of $\epsilon$ it can be easily seen that Eq. (\ref{eom2})
is actually not independent of (\ref{eom1})].

As noted in Ref.\cite{vi91}, the effect of friction is to make the
replacement
$$
H\to H+h
$$
in the equations of motion for a free string in an expanding universe. In
what follows we will consider
$$
a(\tau)\sim \tau^{\alpha}
$$
($\alpha=1$ in the radiation dominated era and $\alpha=2$ in the
matter era). Since $T\sim a^{-1}$, the friction term $h(\tau)$ dominates at
early times, while the expansion term $H(\tau)$ will dominate at late
times. It is clear that friction will be unimportant for $\tau>>\tau_*$,
where $\tau_*$ is the time at which both terms are equal
$$
h(\tau_*)=H(\tau_*).
$$

In terms of cosmological time $t$ defined by $dt=a(\tau)d\tau$ and using
Einstein's equations in the radiation era, $H^2=(8\pi^3 G/90){\cal N}T^4$,
one finds \cite{vi91}
\beq
t_*= A (G\mu)^{-2} t_p,
\label{t*1}
\eeq
(for $G\mu\gesim (A t_p/t_{eq})^{1/2}$).
Here
$t_{eq}$ is the time of equal matter and radiation densities and
$A=[90\beta^{4/3}/32\pi^3{\cal N}]^{3/2}$, with ${\cal N}$ the effective
number of massless degrees of freedom. The coefficient $A$ can be rather
small. Taking $\sin^2\pi\nu_a\sim1/2$ in the expression for $\beta$ one has
$$
A\sim 4\times10^{-4}{\cal N}^{1/2}.
$$
This estimate should actually be taken as an upper bound for $A$, since
some of the particle species may not interact with the string.

If the strings are so light that
$G\mu\lesim (A t_p/t_{eq})^{1/2}$ then
$t_*>t_{eq}$ and (\ref{t*1}) is not valid. Then we have to take into account
that $T\sim t^{-2/3}$ in the matter era and we have
\beq
t_*\approx A^{1/2}(G\mu)^{-1}\left({t_{eq}\over t_p}\right)^{1/2}t_p.
\label{t*2}
\eeq

\section{Linearized perturbations on a straight string}

Consider a straight string at rest, with trajectory $\vec x=(\xi,0,0)$
and introduce a small displacement in the plane transverse to the string,
$$
\delta \vec x= e^{ik\xi}\vec y(\tau)=e^{ik\xi}(0,y_2(\tau),y_3(\tau)).
$$
Substituting in (\ref{eom1}) and keeping only terms linear in $\delta \vec
x$ we have
\beq
\ddot{\vec y}+2\alpha\left[{1\over
\tau}+{\tau_*^{2\alpha-1}\over\tau^{2\alpha}}
\right]\dot{\vec y}+k^2\vec y=0,
\label{linear}
\eeq
where, as mentioned before, $\alpha=1$ in the radiation era and $\alpha=2$
in the matter era.

The case without friction corresponds to
taking $\tau_*=0$. Then
(\ref{linear}) is just the Modified Bessel equation. The corresponding
solution that is well
behaved as $\tau\to 0$ is given by
$$
y=y_0\  2^{\nu}\Gamma(\nu+1)(k\tau)^{-\nu}J_{\nu}(k\tau),
$$
where $\nu=\alpha-1/2$. At early times, $\tau<<k^{-1}$, the solution tends
to a constant,
\beq
y\approx y_0,\quad\quad(\tau<<k^{-1})\label{nofriction0}
\eeq
whereas at late times, $\tau>>k^{-1}$, the solution is
oscillatory
\beq
y\approx y_0\ (2\alpha-1)!!{a(k^{-1})\over a(\tau)}\cos(k\tau-{\alpha\pi\over
2}).\quad(\tau>>k^{-1})\label{nofriction}
\eeq
Modes with wavelength larger than the horizon size ($\tau<<k^{-1}$) are
frozen in, and their physical amplitude $y_{phys}\equiv a(\tau)y$ is
conformally stretched by the expansion. Once the wavelength of a mode falls
within the horizon ($\tau>>k^{-1}$), the perturbation starts oscillating
with constant physical amplitude \cite{vi85}.

When friction is included ($\tau_*\neq 0$), equation (\ref{linear}) can
only be solved in the high frequency approximation ($k>>\tau_*^{-1})$. This
is actually the interesting regime, in which friction plays a role. In the
opposite limit ($k<<\tau_*^{-1}$), we just saw that perturbations are frozen
in up to a time $\tau\sim k^{-1}>>\tau_*$ due to the expansion term
$2\alpha/\tau$ alone, and by the time
they start oscillating the friction term has become negligible.

To solve for the case $k>>\tau_*^{-1}$ we introduce the variable
\beq
\Psi\equiv\tau^{\alpha}\exp\left[{-{\alpha\over2\alpha-1}
\left({\tau_*\over\tau}\right)^
{2\alpha-1}}\right]\ y.\label{Y}
\eeq
Substituting in (\ref{linear}),
one finds that $\Psi$ satisfies the Schr\"odinger equation
\beq
-\ddot \Psi+V(\tau)\Psi=k^2 \Psi,\label{schrodinger}
\eeq
with potential
$$
V\equiv{\alpha\over\tau^2}\left[(\alpha-1)+
\alpha\left({\tau_*\over\tau}\right)^{4\alpha-2}\right].
$$
Now Eq.(\ref{schrodinger}) can be solved in the WKB approximation.
This is done in Appendix A.

With the boundary condition that the perturbations are frozen in with the
expansion at early times, the solution is
\beq
y(\tau)\approx y_0\quad\quad(\tau<<\tau_k) \label{y1}
\eeq
\beq
y(\tau)\approx 2y_0{a(\tau_k)\over a(\tau)}
e^{W(\tau/\tau_*)}
e^{-\gamma k\tau_k}\cos(k\tau+\phi_k),\quad(\tau>>\tau_k).\label{y11}
\eeq
Here
$${W(\tau/\tau_*)}=
{{\alpha\over 2\alpha-1}\left({\tau_*\over\tau}\right)^{2\alpha-1}}
$$
$$
\gamma={1\over 4\alpha-2}B\left({1+2\alpha\over 4\alpha},{1\over 2}\right),
$$
where $B$ is Euler's Beta function and $\tau_k$ is the classical turning
point of the corresponding Schrodinger problem $V(\tau_k)=k^2$. The phase
$\phi_k$ is
just a constant.
In radiation dominated universe
$$
\tau_k=(k^{-1}\tau_*)^{1/2},
$$
$$
\gamma\approx 1.2
$$
In general
\beq
\tau_k\approx[\alpha k^{-1}\tau_*^{2\alpha-1}]^{1/2\alpha}.\label{turningp}
\eeq
It can be checked that the conditions for the validity of
the WKB approximation are satisfied provided that $k>>\tau_*^{-1}$ and
$\alpha>1/2$.

Comparing (\ref{y1}) with (\ref{nofriction}) we can summarize as follows.
Without friction perturbations of wave number $k$ are conformally stretched
up to a time $\tau\sim k^{-1}$, after which they start oscillating with
constant physical amplitude
\beq
y_{phys}\approx (2\alpha-1)!! a(k^{-1}) y_0.\label{ynofriction}
\eeq

With friction included, short wavelength perturbations $k>>\tau_*^{-1}$ are
conformally stretched up to a time $\tau_k$ given by (\ref{turningp}), hence
the factor $a(\tau_k)$ in (\ref{y11}). Since $\tau_k>k^{-1}$, friction
contributes to increase the amplitude of the perturbation by a factor
$(\tau_kk)^{\alpha}$, from time $\tau\sim k^{-1}$ up to time $\tau_k$.
After $\tau_k$, the perturbation starts oscillating and losing energy to
friction. This is represented by the two exponential factors in (\ref{y1}).
By the time $\tau\sim \tau_*$ the amplitude of the perturbations has been
damped by a factor $exp(-\gamma k\tau_k)$. For $\tau>>\tau_*$ the modes do
not lose any more energy to friction, the first exponential in (\ref{y11})
has reached its asymptotic value of unity and the string oscillates with
constant physical amplitude
\beq
y_{phys}\approx2\cdot a(\tau_k)e^{-\gamma k\tau_k}y_0.\label{yfriction}
\eeq

As expected, friction plays no role after $\tau_*$. Also, the exponential
suppression in (\ref{yfriction}) becomes less and less dramatic as $k$
approaches $\tau_*$ (i.e. $k^{-1}\sim \tau_k\sim \tau_*$), in
agreement with the intuitive expectation that friction does not
affect wavelengths of
co-moving size comparable or larger than $\tau_*$.

As mentioned before, the WKB approximation is only valid for $\alpha>1/2$.
The case $\alpha\leq 1/2$ is markedly different because friction never
`switches off'. As a result, the amplitude of perturbations is eventually
damped to zero for all wavelengths. The simplest example is the flat space
case, ($H=0$, $h=constant$.) In this case the equation for $y$ is that of a
damped harmonic oscillator and all excitations disappear after a
characteristic lifetime which is given by $\tau\sim h^{-1}$ for $k>>h$, and
by $\tau\sim 2hk^{-2}$ for $k<<h$ (taking $a=1$).

\section{Circular loops}

A circular loop can be parametrized as
\beq
\vec x(\tau,\xi)=R(\tau)\cdot(\cos(\xi/R_0),\sin(\xi/R_0),0),
\label{circular}
\eeq
where $R_0$ is a constant and $\xi\in[0,2\pi R_0]$. The equations of motion
(\ref{eom1}) and (\ref{eom2}) read (see Appendix B)
\beq
\ddot R+2a(H+h)(1-\dot R^2)\dot R+{(1-\dot R^2)\over R}=0, \label{circ1}
\eeq
\beq
\dot\epsilon+2a(H+h)\dot R^2\epsilon=0,\label{circ2}
\eeq
with
\beq
\epsilon={R\over R_0(1-\dot R^2)^{1/2}}.\label{epsilon}
\eeq

In the case of flat space and no friction, $H=h=0$, $\epsilon$ is a
constant, so Eq.(\ref{eom1}) is linear and one  finds the well known
oscillatory solution
\beq
R(t)=R_0 \cos(t/R_0).\label{ring}
\eeq
Note that, as a consequence of exact circular symmetry, this solution
collapses to a point. Of course the same will happen when we include
friction and expansion. In reality, when the string shrinks to a point
it would form a black hole \cite{vi81,ha89}.
We will consider this process in more detail in Section
\ref{blackhole}, but for now we shall ignore it, and continue the
solutions beyond the singular points. The reason is that here
we are interested
in the energy loss of a loop due to friction. Although we consider the
circular loop for simplicity, similar results would apply to nearly
circular loops which do not shrink to a point.

In the generic case $H+h\neq 0$ we cannot find
exact analytic solutions, so we have to use numerical solutions and
analytic approximations.

\subsection*{a-Loops with friction but no expansion}

It is instructive to start by considering the case $H=0$
and $h=const.\neq 0$.
Consider a loop initially at rest and whose initial radius $R_0$ is
sufficiently large. It is clear that as long as $aR>>h^{-1}$ the motion of
the loop will be overdamped, with characteristic velocity
$$\dot R\approx(2ahR)^{-1}$$
[i.e. we neglect the $\ddot R$ term in (\ref{circ1})]. That this is a good
approximation can be checked by computing $\ddot R$ from the previous equation
and comparing it to the other terms in  (\ref{circ2}).
On the contrary, for $aR<<h^{-1}$ the damping term can be neglected and the
string undergoes a relativistic collapse similar to the frictionless one.
This suggests that the energy of the string at the moment of its first
collapse will be independent of $R_0$ (for $R_0>>h^{-1}$), and will be
roughly equal to the energy of a loop of physical size $h^{-1}$,
$$
E\sim \mu h^{-1}.
$$

This estimate can be made rigorous by using a simple scaling argument.
Introducing dimensionless variables $p\equiv\dot R$, $u\equiv 2haR$ and
$v\equiv 2haR_0\epsilon$, the differential equations (\ref{circ1}) and
(\ref{circ2}) can be reduced to
$$
{dp\over du}=\left[{(1-p^2)^{1/2}\over pv}+(1-p^2)\right],
$$
$$
{dy\over du}=-pv,
$$
where there is no reference to time or to $h$. The initial conditions
$R=R_0$ and $\dot R=0$ become
$$
v(u_0)=u_0, \quad p_(u_0)=0,
$$
where $u_0\equiv 2ha R_0$. The energy of the string at the moment of first
collapse is
\beq
E=2\pi\mu a R_0\epsilon=\chi\cdot\pi\mu h^{-1}.\label{energy1}
\eeq
The coefficient $\chi\equiv v(u=0)$
can be found by numerical integration.
Although $v(u=0)$ depends on $u_0$, the result rapidly saturates to a
constant
$$
\chi\approx .57
$$
for $u_0$ larger than 1.

After the first collapse the loop will undergo a series of oscillations,
losing energy to friction in each one of them. Let us estimate this energy
loss. Since the behaviour of energy is controlled by Eq. (\ref{circ2}) we
will be interested in the quantity $<\dot R^2>$, where the brackets denote
the temporal average between two consecutive collapses of the loop. We have
$$
<\dot R^2>=-<R\ddot R>=<(1-\dot R^2)[1+2ahR\dot R]>=
$$
$$
1-<\dot R^2>-2ah<R\dot R^3>,
$$
or $<\dot R^2>=(1-2ah<R\dot R^3>)/2$, where in the first step we have
integrated by parts and in the second we have used the equation of motion
(\ref{circ1}). Repeating similar steps for the calculation of $<R\dot R^3>$
one can generate a perturbative expansion in powers of $ahR$. The first
terms are
$$
<\dot R^2>={1\over 2}-{1\over 8}a^2h^2<R^2>+0(a^4h^4R^4).
$$
{}From (\ref{energy1}) it is clear that
after the first collapse $ahR\lesim .3$, so the second term is a very small
correction (at most of order $10^{-2}$) to the first and we have
$$
<\dot R^2>\approx 1/2.
$$
{}From (\ref{circ2}) the fraction of the energy lost in
between two consecutive collapses is given by $\Delta E/E\approx -2
<\dot R^2> h T$, where $T$ is the time spent in the oscillation. Therefore
 $E\propto\exp(-ht)$ and
the loop will disappear in a characteristic lifetime of order $\sim h^{-1}$
after the first collapse.

\subsection*{b- Loops with friction and expansion}

The case {\em without} friction in expanding universe has been
previously studied
in the literature \cite{vi85}. Loops are conformally stretched by the
expansion until they cross the horizon. This happens at time
$t_c$ defined by
$$
H^{-1}(t_c)=a(t_c)R_0\equiv r_c.
$$
Here $r_c$ is the physical radius of the loop at horizon crossing. After
that, the loop oscillates with constant physical amplitude. This
behaviour is illustrated in Fig 1a, which corresponds to the numerical
solution of (\ref {circ1}) with $h=0$ and for a radiation dominated universe
($\alpha=1$). The result is plotted in terms of dimensionless cosmological
time $t/t_c$. The upper line represents $E/E_c$, where $E$ is the
energy of the loop and $E_c=2\pi\mu r_c$. In the same figure, we plot
$r/r_c$, where $r=aR$ is the
physical radius, and $\dot R$, the velocity of the string with respect to
the cosmological fluid.

With friction included, for loops such that $r_c>>t_*$, the evolution is
not much different from the one just described. The reason is that by the
time these loops cross the horizon and start oscillating, we already have
$h<<H$.

The effect will be important for loops that cross the horizon well
before $t_*$ (i.e. $r_c<<t_*$). In Fig. 1b we plot the time evolution of a
loop with $r_c=10^{-3} t_*$, in radiation dominated era (same conventions
as in Fig. 1a). Not surprisingly, friction delays the time at which the
loop first collapses. Initially, this increases the energy of the loop,
since it is stretched up to a radius much larger than $t_c$. Later, as the
loop shrinks, it loses energy to friction. It is interesting to observe
that both effects roughly compensate each other, in the sense that at the
time of first collapse, $t_f$ we have $E(t_f)\approx E_c$, just like in the
frictionless case. After $t_f$ the loop keeps losing energy during each
oscillation, up to a time $t_*$, when friction switches off. From that time
on, the loop oscillates with constant physical amplitude.

It is possible to give an approximate analytical description of this
evolution. At early times the motion of the loop is overdamped. Neglecting
the $\ddot R$ term in (\ref{circ1}) we have
\beq
R\dot R\approx {-1\over 2a(H+h)}.
\label{overcirc}
\eeq
Before $t_*$, $H$ can be neglected and we can integrate (\ref{overcirc}) to
find
\beq
R^2=R_0^2-{1\over\alpha(2\alpha+1)}{\tau^{2\alpha+1}\over\tau_*^{2\alpha-1}}.
\label{R}
\eeq
The loop will first collapse at a value of conformal time which is of order
\beq
\tau\sim
\tau_f\equiv[\alpha(2\alpha+1)\tau_*^{2\alpha-1}R_0^2]^{1/(2\alpha+1)}.
\label{tf}
\eeq

Actually, Eq.(\ref{R}) is only valid as long as $\dot
R<<1$. After that, the $\ddot R$ term in (\ref{circ1}) becomes comparable
to the others. Since $R_0<<\tau_f$, the effective friction coefficient, $ah$,
will not appreciably change during the relativistic collapse
(which occurs at a conformal time close to $\tau_f$). From
(\ref{energy1}), the energy of the loop at the moment of first collapse is
$E_f=\approx .57\pi\mu/h(\tau_f).$
This has to be compared to $E_c=2\pi\mu r_c\approx 2\pi\mu R_0a(R_0)$.
Using (\ref{tf}) for $\tau_f$ we have
$$
{E_f\over E_c}\approx\left({\tau_*\over R_0}\right)^{2\alpha^2-3\alpha+1\over
2\alpha+1}.
$$
In the radiation era ($\alpha=1$), we have $E_f\approx E_c$ independently
of $R_0$, in good agreement with the numerical results.
In the matter dominated era, for $R_0<<\tau_*$ we have $E_f>>E_c$, so by
the time $\tau_f$ loops have actually more energy than they would have had
in the absence of friction.

After $t_f$, the radius of the loop is much smaller than the horizon and
the quantity $2a(H+h)$ is slowly varying compared with the period of
oscillation. From the arguments of the previous subsection
we have $<\dot R^2>\approx1/2$, and from (\ref{circ2}) the change in
$\epsilon$ during one oscillation is $\Delta
\epsilon/\epsilon=-a(H+h)\Delta\tau$. Since $\Delta\epsilon<<\epsilon$, one
can take the continuous limit:
$$
{\dot\epsilon\over\epsilon}\approx
-{\dot a\over a}-{\alpha\tau_*^{2\alpha-1}\over \tau^{2\alpha}},
$$
and since $E\propto a\epsilon$, the energy of the loop is given by
\beq
E\approx E_f\
\exp\left[-{\int^{\tau}_{\tau_f}\alpha{\tau_*^{2\alpha-1}
\over \tau^{2\alpha}}}d\tau\right].\label{enas}
\eeq
In the case without friction ($\tau_*=0$) we have $E\approx constant$, in
agreement with the numerical results (Fig. 1a). With friction included, the
energy drops until time $\tau_*$, after which loops oscillate with constant
energy.

\section{Perturbations on circular loops}

Let $R(\tau)$ be a solution of (\ref{circ1}), representing the evolution
of a circular loop. A perturbed loop can be parametrized as
\beq
\rho=R(\tau)+y^{\rho}(\tau,\theta),\label{cyli}
\eeq
$$
z=y^z(\tau,\theta).
$$
Here $(\rho,\theta,z)$ are co-moving cylindrical coordinates, in which the
metric takes the form
$ds^2=a^2(\tau)(-d\tau^2+d\rho^2+\rho^2d\theta^2+dz^2)$, $y^{\rho}$ is
a radial perturbation and $y^z$ is a perturbation transverse to the
plane of the loop.

It is straightforward to write (\ref{eom1}) in cylindrical coordinates and
then substitute (\ref{cyli}) to find the linearized equations for the
perturbations. This is done in Appendix B. Decomposing
$y^{\rho}$ as a sum over modes
\beq
y^{\rho}=\sum_{L=2}^{\infty}
[y^{\rho,+}_L \sin(L\theta)+y^{\rho,-}_L \cos(L\theta)]\label{degeranis0}
\eeq
and similarly for $y^z$, the resulting equations are
\beq
\ddot y^{\rho}_L+2a(H+h)(1-3\dot R^2)\dot y^{\rho}_L-
2{\dot R\over R}\dot y^{\rho}_L+{L^2-1\over
R_0^2\epsilon^2} y^{\rho}_L=0,\label{degeranis1}
\eeq
\beq
\ddot y^z_L+2a(H+h)(1-\dot
R^2)\dot y^z_L+{L^2\over R_0^2\epsilon^2} y^z_L=0.
\label{degeranis2}
\eeq
where we have omitted the index $+$ or $-$.
The sum in (\ref{degeranis0}) starts at $L=2$
rather than at $L=0$
because it is easy to see that
to linear order the $L=0$ and $L=1$ modes do not correspond to deformations
from the circular shape but rather to small translations and rotations of a
circular loop \cite{gavi93}.

Since the function $R(\tau)$ is not known analytically, Eqs.(\ref{degeranis1})
and (\ref{degeranis2}) have to be solved numerically. However, before we do
that, the behaviour of the perturbations can be guessed from the results
of the previous sections.

Let us first consider the case without friction.
Up to the time $t_c$ the loop is conformally stretched by the expansion and
the perturbations will approximately behave as perturbations on a straight
string. Ignoring oscillatory cosine factors, the amplitude of the
perturbations at this time will be
$
y_L(t_c)\approx y_{0,L} a(R_0/L)/a(\tau)
$, where $y_{0,L}$ is the initial perturbation.
Here we have used (\ref{nofriction}) making the substitution $k\to
L/R_0$.
After $t_c$, the loop comes
within the cosmological horizon and starts collapsing. A loop on subhorizon
scales behaves approximately as it would behave in flat space. The theory
of perturbations on a circular loop collapsing in flat space was solved in
\cite{gavi93} (see Appendix A of that reference). It was shown that the
amplitude of transverse perturbations stays constant during the collapse,
whereas the amplitude of radial perturbations shrinks by a factor of $L$
as the loop shrinks to $r<<r_c$. Therefore by that time we have
\beq
y_{phys,L}^{\rho}\equiv a(\tau)y_L^{\rho}\approx y_{0,L}^{\rho}
{a(R_0/L)\over L},\label{she1}
\eeq
\beq
y_{phys,L}^z\equiv a(\tau)y_L^z\approx y_{0,L}^z\ a(R_0/L).
\label{she2}
\eeq

To check the validity of these approximations one has to numerically solve
Eqs.(\ref{degeranis1}) and (\ref{degeranis2}), whith $h=0$. We have done
that for different values of $L$, in the radiation dominated universe
and using the boundary condition that perturbations are initially at rest.
The result is plotted in Fig. 2 as a function of wave number $L$. The
circles denote the ratio $y_{phys,L}^{\rho}$ divided by the r.h.s. of equation
(\ref{she1}) at the time when the loop first collapses. Similarly, the
crosses denote the ratio of $y_{phys,L}^z$ to the r.h.s. of (\ref{she2}).
Ignoring the ``valleys'' in Fig. 2, these ratios are of order 1, which
means that Eqs. (\ref{she1}) and (\ref{she2}) give a very good estimate.
The valleys in Fig. 2 are due to oscillatory behaviour of the
perturbations, which we have ignored in our argument.

The effects of friction can be introduced along similar lines, and they
will only be important for co-moving wavelengths $(R_0/L)<\tau_*$. For such
wavelengths the r.h.s. of equations (\ref{she1}) and (\ref{she2}) has to be
corrected by a factor of
\beq
{2\over(2\alpha-1)!!}(k\tau_k)^{\alpha}\exp(-\gamma k\tau_k).
\label{exponential}
\eeq
This is obtained comparing (\ref{ynofriction}) with (\ref{yfriction}),
where now $k$ is given by $L/R_0$. Again, one can check the accuracy of this
approximation by solving the equations of motion for the perturbations, now
with $h\neq 0$. The results for the case of a radiation dominated universe
are plotted in Fig. 3a for $R_0= \tau_*$, and in Fig. 3b for $R_0= 5 \tau_*$.
The circles and crosses denote the same quantities as in Fig. 2. It is seen
that the suppression factor (\ref{exponential}), depicted as a solid line,
gives the right answer to very good approximation.

The above considerations apply only to loops with $R_0\gesim \tau_*$. For
$R_0<<\tau_*$ the perturbations do not have time to be damped before the
loop starts shrinking, so the factor (\ref{exponential}) actually
overestimates the effect of friction. We shall consider this in more detail
in the next section.

\section{Black hole formation}
\label{blackhole}

Let $r_c$ be the physical radius of a circular loop at horizon crossing. The
mass of this loop is $2\pi r_c\mu$ and the Schwarzschild radius
corresponding to this mass is
$$
r_s=4\pi G\mu r_c.
$$
As the loop shrinks under its tension, its rest mass is converted into
kinetic energy so that the total energy of the loop remains constant
(neglecting friction and gravitational radiation for the moment).
If a loop is exactly circular then it will eventually shrink
to a size smaller than $r_s$ and form a black hole \cite{vi81,ha90}.
(This is only true for strings that form as a result of gauge symmetry
breaking. Strings which form as a result of global symmetry breaking
would radiate all of their energy in the form of Goldstone bososns before
they shrink to the size of their Schwarzschild radius \cite{fova92}).

If the loop is not exactly circular it will still form a black hole
provided that the size of the perturbations is sufficiently small. Let
$\tau_s$ denote the time when the unperturbed loop would shrink to
the size of
its Schwarzschild radius. It is clear that if
\beq
|a(\tau_s)\vec y(\tau_s)|<r_s\  (\equiv a(\tau_s) R_s)
\label{llom}
\eeq
then a black hole will still form.

{}From the analysis of the previous section one could argue that loops of
string with $R_0<< \tau_*$ could easily form black holes. The
argument is that since friction exponentially
suppresses wiggles on scales smaller than $\tau_*$, these loops
would be circular to very good approximation. This would result in the
copious production of black holes with masses up to
\beq
M\sim\mu t_*\sim A(G\mu)^{-1}m_{pl},\label{dial}
\eeq
where $m_{pl}$ is the Planck mass. There are observational upper
bounds on the density of black holes of masses  $M\gesim
10^{14} m_{pl}$, which would be evaporating at the time of nucleosynthesis
or later (see \cite{BCL92} and references therein). Then, from
(\ref{dial}), one would be able to put constraints on topologically
stable cosmic strings of very low tension,
\beq
G\mu<10^{-14}.\label{low}
\eeq

However, this argument needs to be refined, since for $R_0<<\tau_*$ one
cannot simply use the exponential suppression factor (\ref{exponential})
to estimate the effect of friction.
This is because these loops collapse well before $\tau_*$ and friction has
not had enough time to damp the perturbations. Therefore, it is important
to study the behaviour of perturbations on loops with $R_0<<\tau_*$ in more
detail.

At very early times, both the loop and the perturbation will be overdamped.
Neglecting second derivatives in (\ref{degeranis2}) we have
\beq
{\dot y^z\over y^z}=-{L^2\over R^2}{1\over
2a(H+h)}.\label{degenerats}
\eeq
Using (\ref{overcirc}) we have the interesting relation
\beq
y^z(t)=y^z(t_0)\left({R(t)\over R_0}\right)^{L^2},
\label{interesting}
\eeq
so the perturbations shrink faster than the radius of the loop,
the relative perturbation decreasing as the coordinate radius shrinks
from its initial value $R_0$. Following similar steps we also have
\beq
y^{\rho}(t)=y^{\rho}(t_0)\left({R\over R_0}\right)^{L^2-1}.
\label{interesting2}
\eeq
To find the
limit of applicability of the overdamped approximation, one can calculate
$\ddot R$ and $\ddot y^z$ from (\ref{overcirc}) and (\ref{degenerats})
and compare them to the damping term in equations (\ref{circ1}) and
(\ref{degeranis2}). One readily finds that (\ref{overcirc}) is valid for
$\dot R<<1$, while (\ref{interesting}) is valid for $\dot R<<L^{-1}$.
The time at which $\dot R\sim L^{-1}$ coincides also
with the time at which the relevant physical scale comes within the
effective horizon $h^{-1}$
and starts oscillating.

Once a perturbation starts oscillating it is damped very efficiently,
so the perturbations which will be more difficult to eliminate are
those with $L=2$, which are overdamped up to the time when the loop becomes
relativistic, $\dot R\sim 1$. From (\ref{overcirc}) this happens when
$R\sim[ah]^{-1}$, at a time of order $\tau_f$, given by (\ref{tf}). That
is, the loop becomes relativistic when
\beq
{R\over R_0}\sim{1\over a(\tau_f)h(\tau_f)R_0}\sim\left({R_0 \over \tau_*}
\right)^{1/3},
\label{fiu}
\eeq
where in the last step we have restricted attention to loops which are
collapsing in the radiation era.
After that, during the relativistic collapse, the loop is within the
effective horizon and it can be seen that perturbations behave very much
like they would on a circular loop in flat space. That is, as the loop
shrinks the amplitude of transverse perturbatins stays constant whereas
radial perturbations only shrink by a factor of $L$ \cite{gavi93}.

As a result, all the supression in the lowest modes $L=2$ comes from the
overdamped regime. From (\ref{interesting2}) with $(R/R_0)$ given by
(\ref{fiu}), and using that $r_s\sim G\mu h^{-1}(\tau_f)$ we obtain
that at the time $t_s$ when the unperturbed loop crosses its Schwarzschild
radius
\beq
{y^{\rho}_2(t_s)\over R_s}\sim
{y^{\rho}_2(t_0)\over R_0}(G\mu)^{-1}\left({R_0\over \tau_*}\right)^{2/3}
\sim
{y^{\rho}_2(t_0)\over R_0}(G\mu)^{-1}\left({r_c\over t_*}\right)^{1/3}.
\label{directly}
\eeq
For strings formed at a phase transition, the initial value of the
relative perturbation $y^{\rho}_2(t_0)/R_0$ can be of order one. Then, in
order to satisfy (\ref{llom}) we need
$$
r_c\lesim (G\mu)^3t_*.
$$
However, if this condition is met, the energy of the loop at the moment of
first collapse is $E\sim \mu r_c\lesim (G\mu)^2 m_{Pl}$, much smaller than
the planck mass, which simply means that a black hole will not form.

Therefore friction by itself is not sufficient to ensure the formation of a
black hole if we start from an arbitrarily wiggly loop: the lowest modes
$L=2$ are not sufficiently damped.
This was not obvious a priory
and it is in contrast with what would happen if the universe were not
expanding. In that case $h$ is a constant and from
(\ref{interesting}) and (\ref{interesting2}) all loops whose initial size
$aR_0$ is much larger than $(G\mu)^{-1/2}h^{-1}$ would shrink to form black
holes of mass $M\sim \mu h^{-1}$.

Even in an expanding universe,
some of the loops produced at the phase transition
might just happen to be circular enough initially that they would
form black holes, even without friction. With friction included, the
number of
loops that will form black holes is larger. The question is then what
fraction of the ensemble will go into black holes or, in other words, what is
the probability of black hole formation ${\cal P}_{bh}$.

Before we try to answer this question
we should know how ${\cal P}_{bh}$ is constrained from
cosmological observations. Assuming a scale invariant distribution of loops
with number density at horizon crossing given by \cite{vi85}
$dn(r_c)=\nu r_c^{-4} dr_c$, where $\nu$ is a parameter of order one or
smaller, and given that these loops form black holes of
mass $M=2\pi\mu r_c$ with probability ${\cal P}_{bh}$, the number density
of black holes is
\beq
dn(M)=m_{pl}^3\nu(2\pi G\mu)^{3/2}{\cal P}_{bh}
\left({m_{pl}\over M}\right)^{5/2}\left({t_p\over t}\right)^{3/2}
d\left({M\over m_{pl}}\right).\label{mbh}
\eeq
Here we have included a factor of $(r_c/t)^{3/2}$ in the distribution
to account for the dilution of the strings (or black holes) by the expansion
of the universe.
Black holes of mass $M\sim 10^{10}-10^{11}g$ would evaporate during
cosmological nucleosynthesis, producing high energy particles which would
deplete the deuterium and helium when standard nucleosynthesis has almost
concluded. This process has been studied in Ref.\cite{VDN78}, using a
distribution of black holes of the form (\ref{mbh}) with the numerical
coefficient left as a free parameter. The bound that these authors obtained
can be translated into
\beq
(2\pi)^{3/2}\nu (G\mu)^{3/2}{\cal P}_{bh}<10^{-26},\label{deut}
\eeq
(assuming that the present abundance of deuterium is of cosmological
origin).

Note that if ${\cal P}_{bh}\sim 1$, then we would be in conflict with this
constraint for $G\mu\gesim 10^{-18}$. It is then important to estimate ${\cal
P}_{bh}$. An upper bound can be obtained directly from Eq.(\ref{directly}).
The probability that $y^{\rho}_2(t_s)<R_s$ is equal to the probability that
$$
y^{\rho}_2(t_0)<(G\mu)(t_*/r_c)^{1/3}R_0.
$$
Treating $y^{\rho}_2(t_0)$ as a random variable normally distributed, with
r.m.s. amplitude of order $R_0$, and noting that for $L=2$ there are two
independent modes [the + and the $-$ modes
in Eq.(\ref{degeranis0})], that probability is bounded by
\beq
{\cal P}_{bh}<\left[G\mu\left({t_*\over r_c}\right)^{1/3}\right]^2\sim
(G\mu)^{4/3}\left[{m_{pl}\over M}\right]^{2/3}<10^{-9}(G\mu)^{4/3},
\label{hound}
\eeq
where in the last step we have used $M\gesim 10^{14} m_p$. As a result, the
constraint (\ref{deut}) will always be satisfied for cosmic strings in the
low energy range (\ref{low}) that we were interested in.

What can one say
about heavier strings? Note that ${\cal P}_{bh}$ can actually be much lower
than the r.h.s. of (\ref{hound}), and the bound (\ref{deut})
is likely to be satisfied
even for large values of the string tension $G\mu\sim 10^{-6}$. In this
case the loops that form black holes in the mass range
$10^{14}-10^{20}m_{pl}$
cross the horizon much later than $t_*$, so friction only smoothes out
the perturbations of very large $L$.
This does not help very much when we try to form black holes.
The smoothness of
the strings in  this case has to be attributed to other damping mechanisms
such as gravitational radiation.
The probability of black hole formation in the absence of friction has
been estimated by Hawking \cite{ha89} and others \cite{poze91,gavi93},
\beq
{\cal P}_{bh}\sim (G\mu)^p.
\eeq
In the case of a loop formed by $n$ straight segments, Hawking found $p\sim
2n$. This form for ${\cal P}_{bh}$ results from the fact that in order to
form a black hole it is necessary to ``fine tune'' a set of $2n$ angles
with acuracy given by $G\mu$. Typically, $p$ will be of the order of the
number of random variables that parametrize the ensemble of loops, since
for a black hole to form all the parameters have to be fine tuned with an
accuracy given by $G\mu$. Polnarev and Zembowicz \cite{poze91} studied
${\cal P}_{bh}$ for a family of loops containing excitations in the first and
third harmonics in a Fourier expansion of the solutions to the Nambu
equations of motion. They studied a two parameter family and they found
$p\approx 2$ (with some uncertainty due to the arbitrariness in the
definition of a probability distribution in the space of parameters.)
However, $p$ is likely to be larger (and therefore ${\cal P}_{bh}$
smaller) since one need not be restricted to a two parameter family. In
particular, the general solution including the first and third harmonics is
a five parameter family (see e.g. \cite{br90}), and one may expect
$p\sim 5$.

Other observational constraints come from black holes of mass $M\sim
10^{20}m_{pl}$, which would be evaporating at the present time, producing
intense bursts of $\gamma$-rays (see e.g. \cite{maca91}).
This constraint has been studied in \cite{poze91}, whose authors concluded
that if $p\lesim 2$ there is conflict with observations for strings
heavier than $G\mu\gesim 10^{-7}$. On the other hand if $p>4$ there is no
conflict even if $G\mu$ is as large as $10^{-5}$. Including friction does
not modify these results, since for $M\sim 10^{20}m_{pl}$ friction is
important only for $G\mu\lesim 10^{-20}$ [see Eq.(\ref{dial})]. But
for such low values of the tension there is no observable
effect even if ${\cal P}_{bh}=1$ \cite{poze91}.

\section{Strings attached to walls}

In this section we consider perturbations on strings which are
attached to planar domain walls \cite{vi85}.
For simplicity, we shall ignore the expansion of
the universe. Also we shall restrict ourselves to perturbations which lie
on the plane of the wall.

Then we are effectively left with a 2+1 dimensional problem, in which the
string is the boundary between a region of ``false vacuum'' (the wall) with
energy per unit area equal to the surface tension $\sigma$, and a region
of ``true vacuum'' where there is no wall. In the absence of friction, the
equations of motion are \cite{gavi91}
\beq
-\mu\Box x^{\mu}=\sigma n^{\mu}.
\label{eomnof}
\eeq
Here $\Box$ is the covariant d'Alembertian on the
worldsheet of the string and $n^{\mu}$ is the space-like unit vector normal
to the worldsheet, with $n^{\mu}n_{\mu}=1$ and
$n_{\mu}\partial_ax^{\mu}=0$. Our sign convention is that $n^{\mu}$ points
towards the wall. It is easy to see that
equation (\ref{eomnof}) has a solution representing a
straight string which
is constantly accelerating due to the tension of the wall. The wall gradually
disappears
as the string moves forward, its energy going into kinetic
energy of the string \cite{gavi91}. The string
undergoes the so-called hyperbolic motion, asymptotically approaching the
speed of light at late times.

Eq. (\ref{eomnof}) has to be modified to include the force of friction
$F^{\nu}$. To keep things general we take
\beq
F^{\nu}=F(u_{\bot})n^{\nu},
\eeq
without specifying the dependence in the transverse velocity of the fluid
$u_{\bot}=u^{\mu} n_{\mu}$. (Note that $u_{\bot}^{\nu}$ defined in Section 2
is given by $u_{\bot}^{\nu}=u_{\bot}n^{\nu}$.) Then the equations of motion
read
\beq
-\mu\Box x^{\nu}=[\sigma+F(u_{\bot})]n^{\nu}.
\label{eomf}
\eeq

We can study perturbations on any solution $x^{\mu}(\xi^a)$ of (\ref{eomf})
using the covariant formalism of Ref. \cite{gavi91},
suitably modified to include friction.
Since only perturbations which are normal to the string are physically
observable, one need only consider perturbations of the form
$$
\delta x^{\mu}=\phi(\xi)n^{\mu}.
$$
Here $\phi$ represents the proper magnitude of the perturbation, i.e. the
normal displacement as measured by an observer that is moving with
the string.
Multiplying (\ref{eomf}) by
$n_{\nu}$ we can write
\beq
-\mu K^{a}_a=(\sigma+F),
\label{ny}
\eeq
where $K_{ab}=n_{\mu}\nabla_a\partial_b x^{\mu}$ is the extrinsic
curvature
of the worldsheet.
Latin indices are raised and lowered using the worldsheet metric
$$
g_{ab}\equiv \partial_a x^{\mu}\partial_b x_{\mu}.
$$
The linearized equation of motion for the perturbations can be found from
variation of Eq.(\ref{ny})
$$\mu \delta K^a_a=-\delta F,$$
where $\delta$ denotes the variation induced, to linear order in $\phi$, by
the small perturbation $\delta x^{\mu}$. Following \cite{gavi91}, we have
$\delta K^a_a=\Box \phi+K^{ab}K_{ab}\phi.$ On the other hand
$\delta F=F'(u_{\bot})\delta u_{\bot}$, where
$F'=dF/du_{\bot}$. Also, $\delta u_{\bot}=u_{\mu}\delta n^{\mu}$, and the
change in $n^{\mu}$ induced by the perturbation is \cite{gavi91}
$\delta n^{\mu}=-g^{ab}\phi,_b\partial_a x^{\mu}$. Therefore,
the resulting equation of motion for $\phi$ is
\beq
-\Box \phi+M^2 \phi=-{F'\over\mu}u_{\mu}\partial_ax^{\mu}\phi,^a\ ,
\label{reseq}
\eeq
where the ``mass'' is given by
\beq
M^2=-K_{ab}K^{ab}=
\left[{\cal R}-\left({\sigma+F\over \mu}\right)^2\right].
\label{mass}
\eeq
In the last step we have used the Gauss-Codazzi relation
$K_{ab}K^{ab}=(K^a_a)^2-{\cal R}$, where ${\cal R}$ is the intrinsic
curvature scalar on the worldsheet.
The r.h.s. of eq. (\ref{reseq})
acts like a friction term for the perturbation $\phi$ provided that
$F'>0$. But this condition is always met, since it just means that the force of
friction increases with the transverse velocity of the fluid.

In the spirit of Refs.\cite{gavi91,gavi92}, Eq.(\ref{reseq}) can be seen as the
equation of motion for a scalar field $\phi$ which is ``living'' in the
worldsheet of the defect. Now this equation has a different mass than in
the case without friction, and it also has a right hand side in which the
field has a derivative coupling to some external sources (which are
essentially
the `temporal' components of the tangent vectors to the string.)

In the case when the unperturbed string is straight,
lying along the $y$ axis and moving with trajectory $x=x(t)$,
the metric on the worldsheet is
$$
ds^2=g_{ab}d\xi^a d\xi^b=-(1-\dot x^2)dt^2+dy^2=-d\tau^2+dy^2.
$$
Here $t$ is the time in the rest frame of the fluid and $\tau$ is the proper
time of an observer who is moving with the string.
In this case the metric on the worldsheet is flat, the curvature scalar
vanishes and the effective mass $M^2$ is tachyonic. This is also true in the
absence of friction \cite{gavi91}, and it essentially means that modes with
wavelength larger than $|M|^{-1}$ are unstable. Now, including friction,
the difference is that the tachyonic mass ``switches off'' as the string
approaches its terminal velocity $v$. Indeed, the terminal velocity of
the string
is determined by the vanishing of the driving force in (\ref{eomf}), that is
\beq
\sigma+F(u_{\bot})=0,
\label{terminal}
\eeq
but this implies that the mass term
for the perturbations vanishes.

When the straight string
has reached the terminal velocity $v$, equation (\ref{reseq}) reduces to
\beq
{d^2\phi\over d\tau^2}+k^2\phi=-{F'\gamma_v\over\mu}{d\phi\over d\tau},
\label{easy}
\eeq
where $k$ is the wave number of the perturbation, and
$\gamma_v=(1-v^2)^{-1/2}$ is the relativistic factor corresponding to the
terminal velocity. The friction term in the r.h.s. of (\ref{easy}) now
causes the perturbations to decay exponentially with a lifetime which is
easily calculable.
Taking a force of the form (\ref{2}) we have $F'=\beta
T^3$. Noting that $u_{\bot}=n^{\mu}u_{\mu}=-v \gamma_v$, Eq.
(\ref{terminal}) gives
$$
\gamma_v v={\sigma\over \beta T^3}.
$$
In cosmological situations the temperature is always lower than the energy
scale of the wall (typically of order $\sigma^{1/3}$), therefore the
terminal velocity will be relativistic
$$v\sim 1$$
and the gamma factor will
be of order $\gamma_v\sim \sigma T^3$ (ignoring the numerical factor
$\beta$). Then the r.h.s. of (\ref{easy}) reads
$$
{F'\gamma_v \over \mu}\sim {\sigma\over\mu}.
$$
It is then straightforward to see that the perturbations decay with
proper lifetime given by
$\tau\sim {\mu/\sigma}$ for $k>>{\sigma/\mu}$ and $\tau\sim
\sigma\mu^{-1}k^{-2}$ for $k<<\sigma/\mu$.

A more interesting case is that of a string which is at the boundary
of a circular hole that has spontaneously nucleated on a metastable domain
wall (see e.g. Ref.\cite{prvi92}). This process is
the 2+1 dimensional analogue of the
formation of true vacuum bubbles in the problem of false vacuum decay.

Without friction, perturbations on these bubbles (or holes) have been
studied in ref. \cite{gavi91}. In that case, the unperturbed solution was a
circular hole whose radius $R$ expands with constant acceleration,
\beq
R^2-t^2=(2\mu/\sigma)^2. \label{yeye}
\eeq
The effective mass for the perturbations
was $M^2=-\sigma/2\mu^2$, which is tachyonic. Because of the expansion
of the hole, any
perturbation eventually reaches a wavelength larger than $M^{-1}$, at which
point it becomes unstable and starts growing like $\phi\propto R\approx t$
\cite{gavi91}.
{}From an intrinsic point of view, one can say that the string is
unstable, in the sense that $\phi$, (the perturbation measured by a
co-moving observer) grows in time. However, an external observer measures a
perturbation which is Lorentz contracted \cite{gavi91}
$\Delta=\gamma^{-1}\phi$, where here $\gamma=(1-\dot R^2)^{-1/2}$. Since
from (\ref{yeye}) $\gamma\sim R$, we have $\Delta\sim const.$ at
late times. The relative perturbation $\Delta/ R$ decreases and the string
becomes more circular as the hole expands \cite{gavi91}.

Including friction, the main difference will be that the string reaches a
terminal velocity $v$, and the Lorentz contraction factor will go to a
constant at late times. Also, the mass term switches off as the string
approaches the terminal velocity. Let us see, then, what is the fate of the
perturbations when friction is included in the dynamics.
The string worldsheet is given by $x^{\mu}=(t, R\cos\theta,R
sin\theta,0)$, and the metric induced on the worldsheet is
$$
ds^2=-(1-\dot R^2)dt^2+R^2  d\theta^2=-d\tau^2+R^2d\theta^2.
$$
The normal vector is $n_{\mu}=(1-\dot R^2)^{-1/2}(-\dot R,
\cos\theta,\sin\theta,0)$. The extrinsic curvature can be easily calculated
and one finds
\beq
-M^2=K_{ab}K^{ab}=\gamma^2 R^{-2}+\gamma^6(\ddot R)^2.
\label{90}
\eeq
The velocity is bounded by the terminal velocity $\dot R< v$, so at late
times $R\approx vt$ and $\ddot R$ decays faster than $R^{-1}$. Then,
neglecting the $\ddot R$ term in (\ref{90}) the equation of motion for the
perturbations takes the form
\beq
{d^2\phi_L\over d\tau^2}+{F'\gamma_v\over\mu} {d\phi_L\over d\tau}+
{L^2-\gamma_v^2\over R^2}\phi_L =0,
\label{91}
\eeq
with $\phi=\phi_L(\tau)\exp(iL\theta)$.

Note that as the hole expands, the tachyonic mass squared
$-\gamma_v^2 R^{-2}$ ``redshifts'' at the same rate as the wave number term
$L^2 R^{-2}$. One might expect instability for modes with
$L<\gamma_v$, since for these the total effective mass squared is negative.
However, at sufficiently late times we can use the overdamped approximation
and neglect $\ddot \phi$ in (\ref{91}). Then a simple integration shows
that, since $R^{-2}$ decreases faster than $\tau^{-1}$, the
amplitude of the perturbations
asymptotically approaches a constant $\phi\to const.$
{}From an intrinsic point of view the
asymptotic behaviour of the perturbations
is very different from the frictionless case,
in which $\phi$ was growing in time.
However, now $\gamma_v$ is constant and we have
$\Delta=\gamma^{-1}_v\phi \to const.$, so from the point of view of an
external observer which is at rest
the behaviour is similar to that of
the frictionless case. The relative perturbation $\Delta/R$ also goes to
zero, and the loop becomes more and more circular as the hole expands.

\section{Conclusions}
\label{conclusions}

In this paper we have studied the evolution of cosmic strings taking
into account the force of friction. The results are conveniently expressed
in terms of $t_*$, defined as the time at which friction ``switches off''
[equations (\ref{t*1}) and (\ref{t*2})].

For small perturbations around a straight string, a WKB analysis shows that
they are exponentially suppressed if their wavelength crosses the horizon
before the time $t_*$.
Relative to the frictionless case, the amplitude of the perturbations has to
be multiplied by the suppression factor
(\ref{exponential}), which in the case of a radiation dominated universe
($\alpha=1$) reduces to
\beq
4\left({\pi t_*\over \lambda}\right)^{1/2}
\exp\left[-2\gamma\left({\pi t_*\over\lambda}\right)^{1/2}\right].
\label{supre}
\eeq
Here $\lambda$ is the physical wavelength of the perturbation at the time
$t_*$ and $\gamma\approx 1.2$. The suppression of the amplitude of these
perturbations is due to the oscillation of the perturbations
before the time $t_*$. After $t_*$ the perturbations
oscillate with constant physical amplitude.
Perturbations whose wavelength crosses the horizon after $t_*$ are
practically unaffected by friction.

Similarly, for circular loops we should distinguish between large loops
with $r_c>>t_*$, and small loops with $r_c<<t_*$. Here $r_c$ is the radius
of the loop at horizon crossing. Large loops are unaffected by friction,
after they cross the horizon they start oscillating with constant physical
amplitude $r_c$. Small loops do {\em not} start collapsing relativistically
right after they cross
the horizon.
During the non-relativistic evolution, the velocity
at which the string moves with respect to the fluid is given by
$$
\dot R\sim {\mu\over \beta T^3 r},
$$
where $r$ is the physical radius of the loop, T is the temperature and $\beta$
is the numerical factor in (\ref{2}).
Due to the frictional force, these loops are dragged by the expansion more
than they would in the absence of friction,
growing to a size larger than $r_c$.
However as they collapse they lose energy to friction. It is interesting
to note that in a radiation dominated universe both effects approximately
compensate each other, in the sense that at the time when they first
collapse to a point their energy is given by $E_f\approx 2\pi\mu r_c$, just
as in the frictionless case (in matter dominated universe $E_f$ is actually
larger than $2\pi\mu r_c$).
In
subsequent oscillations, the loops will keep losing energy, up
to the time $t_*$. After that they oscillate with constant energy. This
energy can be obtained from (\ref{enas}) in the limit of large times, and
one has $E(t>>t_*)\approx E_f\exp(-2t_*/3r_c)^{1/3}$.

We have also studied perturbations on circular loops. If the loops are
larger than $t_*$, then the perturbations behave in much the same way as
the perturbations on a straight infinite string. Perturbations
whose wavelength crosses
the horizon before $t_*$ are suppressed according to (\ref{supre}), whereas
if it crosses later than $t_*$ they are unaffected by friction.
If the loops are much smaller than $t_*$ then it is not correct to use the
suppression factor (\ref{supre}), essentially because the loops start their
relativistic collapse much before $t_*$. As a result, the perturbations
in the lowest modes are only suppressed
as a power law in $(r_c/t_*)$ [see e.g. (\ref{directly})].

Because the suppression in the relative amplitude of the lowest modes is
only power law, the probability that loops of strings become circular
enough to form black holes is very small (to form a black hole one needs
that the relative amplitude of the perturbations be of order $G\mu$, which is
very small). As a result, the number of black holes of masses
$M\gesim 10^{14}m_{Pl}$ (which would be evaporating at the time of
nucleosynthesis or later) formed by strings which cross the horizon before
$t_*$ is too small to have any observable consequences (for all values of
$\mu$). For loops which cross the horizon later than $t_*$, friction only
eliminates the perturbations of very large wavenumber, so the probability
that they form black holes is not substantially enhanced by friction.

We have also studied perturbations on strings attached to walls. For this
we have used the covariant formalism developed in ref. \cite{gavi91},
generalizing it to include the force of friction. In particular we have
studied the case of a circular hole which spontaneously nucleates on a
metastable domain wall. We find that perturbations on the string that is at
the boundary of the hole are initially unstable, growing at the same rate
as the radius of the expanding hole.
However, as the loop reaches its terminal velocity $v<1$, the instability
switches off and the perturbations
freeze out at a constant amplitude. As a result, the relative perturbation
decreases in time, and the hole becomes increasingly circular as it
expands. This behaviour is similar to the behaviour that one obtains in the
absence of friction \cite{gavi91}. However the mechanism by which
perturbations freeze out is very different. In the former case it is due to
the force of friction, which balances the instability due to the tension of
the wall. In the latter case the freezing occurred for purely kinematical
reasons, since the unperturbed string asymptotically approached the speed
of light.

\section*{Acknowledgements}

M.S. would like to thank the Tufts Institute of Cosmology for hospitality
during the preparation of this work. M.S. was partially supported by the
Directorate-General for Science Research and the development  of the
Commision of the European Communities under contract No. C11-0540-C.
J.G. acknowledges support from the National Science Foundation under Grant No.
PHY-9248784.

\appendix
\section*{Appendix A}
\setcounter{equation}{0}
\renewcommand{\theequation}{A\arabic{equation}}

Let us solve Eq.(\ref{schrodinger}) in the WKB approximation.
The solution `under the barrier' is given by
\beq
\Psi=\sum_{\pm}{C_{\pm}\over\sqrt{p_1}}
\exp\left[{\pm\int^{\tau_k}_{\tau}d\tau'p_1(\tau')}\right],
\quad(\tau<<\tau_k)
\label{wkb1}
\eeq
where $p_1\equiv[V(\tau)-k^2]^{1/2}$ and $\tau_k$ is the `classical' turning
point
$$
V(\tau_k)=k^2.
$$

Since at early times
$V(\tau)\sim\alpha^2\tau_*^{-2}(\tau_*/\tau)^{4\alpha}$,
it is clear that
\beq
I\equiv\int_{\tau}^{\tau_k}d\tau'p_1(\tau') \to {\alpha\over2\alpha-1}
\left({\tau_*\over\tau}\right)^{2\alpha-1}-\Gamma(k,\tau_k)+O(\tau^{2\alpha-1}),
\label{exponent}
\eeq
where the function $\Gamma(k,\tau_k)$, which will be specified below, is
independent of $\tau$.

Since perturbations are frozen in at early times,
the coefficients $C_{\pm}$ are determined by imposing the boundary
condition that the co-moving perturbation $y$ should approach some constant
value $y_0$ as $\tau\to 0$. This immediately requires
$$
C_+=0,
$$
$$
C_-=y_0\sqrt{\alpha\tau_*^{2\alpha-1}}e^{-\Gamma(k,\tau_k)}.
$$

Using the standard connection formulas of the WKB formalism, the solution
in the `classically allowed' region is
\beq
\Psi=2y_0\left({\alpha\tau_*^{2\alpha-1}\over p_2}\right)^{1/2}
e^{-\Gamma(k,\tau_k)}
\cos\left(\int_{\tau_k}^{\tau}d\tau'p_2(\tau')-{\pi\over 4}
\right),\quad(\tau>>\tau_k),\label{wkb2}
\eeq
where $p_2=[k^2-V]^{1/2}$.

The simplest case, which is also the most interesting, is the radiation
dominated universe $\alpha=1$. In this case the  potential assumes a
particularly simple form $V=\tau_*^2\tau^{-4}$, and the turning
point is given by $\tau_k=(\tau_*k^{-1})^{1/2}$. Then,
the exponent $\Gamma(k,\tau_k)$ can be calculated analytically
\beq
\Gamma(k,\tau_k)=-\lim_{\tau\to 0}
\left[\int_{\tau}^{\tau_k} d\tau\left({\tau_*^2\over\tau^4}-k^2\right)^{1/2}
d\tau -\left({\tau_*\over\tau}\right)\right].
\label{Gamma1}
\eeq
Integrating by parts we have
\beq
\Gamma=k\tau_k \lim_{x\to 0}2\int_x^1{dx\over (x^{-4}-1)^{1/2}}=
k\tau_k{1\over 2}B({3\over 2},{1\over 2}),
\eeq
where $B$ is Euler's Beta function.

For $\alpha\neq 1$, the potential is more complicated and the turning
points cannot be given explicitly in terms of $\tau_*$ and $k$. However,
for $k>>\tau_*^{-1}$, the potential under the barrier can be approximated
by the second term and we have
\beq
\tau_k\approx[\alpha k^{-1}\tau_*^{2\alpha-1}]^{1/2\alpha}.\label{turning}
\eeq
Neglecting the first term in $V(\tau)$ and following the same steps as
before we have
$$
\Gamma=k\tau_k
{1\over 4\alpha-2}B\left({1+2\alpha\over 4\alpha},{1\over 2}\right).
$$
Substituting in (\ref{wkb1}), (\ref{wkb2}) and using (\ref{Y}) we have the
solution (\ref{y11}).

\appendix
\section*{Appendix B}
\setcounter{equation}{0}
\renewcommand{\theequation}{B\arabic{equation}}

Here we express the equations of motion for a string with friction in
cylindrical coordinates, and derive the equations for perturbations on a
circular loop.

The FRW metric is written as
$$
ds^2=a^2(\tau)(-d\tau^2+d\rho^2+\rho^2d\theta^2+dz^2).
$$
The nonvanishing Christoffel symbols are
$$
\Gamma^0_{00}=\Gamma^{\theta}_{\theta 0}=\Gamma^z_{z 0}=
\Gamma^{\rho}_{\rho 0}={\dot a\over a},
$$
$$
\Gamma^{\rho}_{\theta\theta}=-\rho;\
\Gamma^{\theta}_{\theta\rho}=\rho^{-1};\
\Gamma^0_{ij}={\dot a\over a}\tilde g_{ij}.
$$
The metric on the string worldsheet is given by
$$
g_{ab}d\xi^ad\xi^b=a^2(\tau)[-(1-v^2)d\tau^2+p^2d\xi^2],
$$
where
$$
v^2\equiv (\dot{\vec x})^2=\tilde g_{ij} \dot x^i \dot x^j,
$$
$$
p^2\equiv (\vec x')^2=\tilde g_{ij} x'^ix'^j.
$$
An overdot denotes derivative with respect to $\tau$, and a prime is
derivative with respect to $\xi$. Here $\tilde g_{ij}$ is the flat
three-dimensional metric in cylindrical coordinates, and the indices $i,j$
run over $\rho,\theta$ and $z$.

Substituting in (\ref{1}), the temporal component yields
$$
\dot \epsilon+ A(\tau) v^2\epsilon=0,
$$
where $A(\tau)=2a(H+h)$ and $\epsilon\equiv p(1-v^2)^{-1/2}$. The spatial
components yield
\beq
\ddot\rho+(1-v^2)A(\tau)\dot\rho-{1\over\epsilon}
\left({\rho'\over\epsilon}\right)'-\rho\left[\dot
\theta^2-{\theta'^2\over\epsilon^2}\right]=0,
\label{rrho}
\eeq
\beq
\ddot \theta+(1-v^2)A(\tau)\dot\theta-{1\over\epsilon}\left(
{\theta'\over\epsilon}\right)'+{2\over\rho}\left[\dot\theta\dot\rho-
{\theta'\rho'\over\epsilon^2}\right]=0,
\label{ttheta}
\eeq
\beq
\ddot z+(1-v^2)A(\tau)\dot
z-{1\over\epsilon}\left({z'\over\epsilon}\right)'=0.
\label{zzeta}
\eeq

For circular loops we can take $z=0$ and
the parameter $\xi$ proportional to
the angular variable, $\xi=R_0\theta$, where $R_0$ is the initial
coordinate radius of the loop. Then $\rho=R(\tau)$, $\rho'=0$,
$\dot\theta=0$ and we obtain (\ref{circ1}) and (\ref{circ2}).

Let us derive the equations for the perturbations on circular loops. Taking
$z=y^z(\tau\theta)<<R$ it is immediate from (\ref{zzeta}) to obtain the
equation for small transverse perturbations (\ref{degeranis2}).
The equation for the radial perturbations requires more work. First we write
the perturbed  solution as $\rho=R(\tau)+\Delta$, and
$\theta=(\xi/R_0)+\delta$,
where $\Delta$ and $\delta$ are small deviations from the unperturbed values.
Substituting in (\ref{rrho}) one finds, to linear order in the perturbations
$$
\ddot R+\ddot\Delta+(1-\dot R^2-2\dot R\dot\Delta)A\dot R\left(1+{\dot\Delta
\over\dot
R}\right)-{\Delta''\over\epsilon^2}+(R+\Delta){1+2\delta'\over\epsilon^2}=0.
$$
Also to linear order
$$
\epsilon^2=\epsilon_0^2\left[1+{2\Delta\over R}+{2\dot R\dot\Delta\over
1-\dot R^2}\right](1+2\delta'),$$
where $\epsilon_0$ is the unperturbed value.
Substituting in the previous expression and using (\ref{circ1}) we find
$$
\ddot\Delta-2{\dot R\over R}\dot\Delta+(1-3\dot R^2) A(\tau)\dot\Delta +
{L^2-1\over\epsilon^2}\Delta=0,
$$
which is the equation of motion for radial perturbations (\ref{degeranis1}).

\section*{Figure captions}
\begin{itemize}
\item {\bf Fig.1a} Evolution of a circular loop without friction in
radiation
dominated universe, as a function of cosmological time $t$. Here $t_c$ is
the time at which the loop crosses the horizon, $r$ is the physical radius
of the loop and $r_c$ is the radius at horizon crossing. The
energy of the loop $E$ rapidly approaches the value $E_c\equiv 2\pi\mu r_c$
after the loop crosses the horizon, and remains approximately constant. We
also plot $\dot R$, the velocity of the string with respect to the
cosmological fluid.
\item{\bf Fig.1b} Evolution of a circular loop with friction in radiation
dominated universe, with $r_c=10^{-3} t_*$. (Same conventions as in Fig.
1a.)
\item{\bf Fig.2} Numerical results for the evolution of perturbations on a
circular loop in radiation dominated universe, ignoring the force of
friction. The results are plotted as a function of wave number $L$. The
circles denote the ratio $y_{phys,L}^{\rho}$ divided by the r.h.s. of
equation (\ref{she1}) at the time when the loop first collapses. Similarly
the crosses denote the ratio of $y_{phys,L}^z$ to the r.h.s. of
(\ref{she2}). Ignoring the oscillatory behaviour, this ratios are of order
one, which means that (\ref{she1}) and (\ref{she2}) give a very
good estimate.

\item{\bf Fig.3a} Numerical results for the evolution of perturbations on
a circular loop in radiation dominated universe, including the force of
friction, for $R_0=\tau_*$. Circles and crosses denote the same quantities
as in Fig. 3. It is seen that the suppression factor (\ref{exponential}),
depicted as a solid line, gives the right answer to very good
approximation.

\item{\bf Fig.3b} Same quantities as in Fig. 3a, for the case
$R_0=5 \tau_*$.

\end{itemize}

\end{document}